\documentclass{aastex6}
\shorttitle{HIMF Variation}
\shortauthors{R. F. Minchin}

\begin{document}
\title{Can the discrepancy between locally and globally derived neutral hydrogen mass functions be explained by a varying value of $M^\star$?}
\author{Robert F. Minchin}
\affil{Arecibo Observatory/USRA \\HC3 Box 53995, Arecibo PR 00612}
\email{rminchin@naic.edu}

\begin{abstract}
I investigate whether it is possible to reconcile the recent ALFALFA observation that the neutral hydrogen (HI) mass function (HIMF) across different galactic densities has the same, non-flat, faint-end slope, with the observations of isolated galaxies and many galaxy groups that show their HIMFs to have flat faint-end slopes. I find that a fairly simple model in which the position of the knee in the mass function of each individual group is allowed to vary is able to account for both of these observations. If this model reflects reality, the ALFALFA result points to an interesting `conspiracy' whereby the differing  group HIMFs always sum up to form global HIMFs with the same faint-end slope in different environments. More generally, this result implies that global environmental HIMFs do not necessarily reflect the HIMFs in individual groups belonging to that environment, and cannot be used to directly measure variations in group-specific HIMFs with environment.
\end{abstract}
\keywords{galaxies: mass function}

\section{Introduction}

The neutral hydrogen (HI) mass function (HIMF) is normally parameterized using a Schechter function:
\begin{equation}
\phi(M) dM = \phi^\star \left(\frac{M}{M^\star}\right)^\alpha e^{-(M/M^\star)} d\left(\frac{M}{M^\star}\right)
\end{equation}
Where $\alpha$ is the faint-end power-law slope, $M^\star$ is the position of the knee, where the function transitions from a power-law to an exponential cutoff, and $\phi^\star$ is the density normalization. 

Recently, \citet{jones2016} measured how the neutral hydrogen (HI) mass function (HIMF) varies with density of galaxies using data from the Arecibo Legacy Fast Arecibo L-band Feed Array (ALFALFA) survey 70\% catalogue. They found that the faint-end slope does not vary significantly between the different densities measured. This differed from the earlier results of \citet{zwaan2005}, who found a faint-end slope that steepened in denser environments using data from the blind HI Parkes All Sky Survey (HIPASS), and \citet{springob2005}, who found a shallower faint-end slope in denser environments using data from the pointed Arecibo General Catalogue (AGC) survey, both using far fewer galaxies. 

\subsection{Tensions between the ALFALFA result and localized observations}

\citeauthor{jones2016} noted that their result was surprising as more localized observations of groups of galaxies (\citealt{freeland2009}; \citealt{kovac2005}; \citealt{pisano2011}; \citealt{verheijen2001}) had found HIMFs with flat ($\alpha \sim -1$) faint-end slopes, and they expected that this would be reflected in their global measurement. 

The situation in groups is complex, as a wide range of faint end slopes have been measured in these environments. Besides the flat or near-flat HIMFs considered by \citeauthor{jones2016}, HIMFs consistent with or steeper than the field ($\alpha = -1.33 \pm 0.02$; \citealt{martin2010}) were found, for example, by \citet{banks1999} in the Centaurus A group using HIPASS data ($\alpha = -1.30 \pm 0.15$) and by \citet{stierwalt2009} in the Leo I group using ALFALFA data ($\alpha = -1.41^{+0.2}_{-0.1}$), while a declining slope was found by \citet{kilborn2009} in southern GEMS groups ($\alpha =0.00 \pm 0.18$). The situation with the lower-density environments of isolated galaxies is simpler, however, and here, again, the \citeauthor{jones2016} results appears to be in contradiction with observations. HIMFs consistent with being flat, and shallower than the field HIMF were found around isolated galaxies by \citet{pisano2003} and in the recent work of \citet{minchin2016} (possibly due in both cases to the tendency of satellites near large galaxies to be gas-poor; \citet{spekkens2014}).

\citet{jones2016} propose that this points to measurement errors in determinations of local HIMFs in galaxy groups due to the limitations of interferometric surveys or, for their nearest-neighbor analysis, that the surface-density of galaxies is independent of group size so that groups are not well separated by this method; here I propose an alternative explanation that unifies the two set of observations without the need to  assume either of these conditions.

\section{Model}

It is known that the luminosity function of isolated galaxies is flat or increasing \citep{verdes2005}, thus isolated galaxies are found at a range of different luminosities and, it is reasonable to assume, different HI masses. This means that the knees in the local HIMFs around isolated galaxies must be found across a range of masses, rather than at a single value -- which could only be the case if all isolated galaxies had approximately the same HI mass. As the global HIMF consists, by definition, of the sum of local HIMFs, and the local HIMF around isolated galaxies must cutoff at the mass of the isolated galaxy, I thus investigate a model where a global HIMF is constructed from local HIMFs with varying values for the cut-off, $M^\star$.

A simple model can be built assuming that $M^\star$ has a cutoff at some value, $M^\star_0$, and is distributed so it is (like the isolated galaxy luminosity function) flat beneath this, so $\phi^\star = \phi^\star_0 (M^\star/M^\star_0)^{-1}$. This can be generalized by assuming that $M^\star$ is distributed below the cut-off according to a power-law with slope $\beta$ (in the flat case, $\beta = -1$), giving the modified Schechter function:
\begin{equation}\label{eqn2}
\phi(M,M^\star) dM = \phi^\star_0 \left(\frac{M^\star}{M^\star_0}\right)^\beta \left(\frac{M}{M^\star}\right)^\alpha e^{-(M/M^\star)} d\left(\frac{M}{M^\star_0}\right)
\end{equation}
In order to examine a continuous distribution it is necessary to integrate Equation \ref{eqn2} with respect to log $M^\star$ up to the limit log $M^\star_0$. Noting that $d(\log M^\star) = M^{\star -1} dM^\star$ and then substituting in $t = M/M^\star$, so $dt = (-t^2/M) dM^\star$. This gives $dM^\star = -t^{-2} M dt$ and thus $d(\log M^\star) = -t^{-2} (M/M^\star) dt = -t^{-1} dt$, with the limits becoming $\infty$ and $M/M^\star_0$, giving:
\begin{eqnarray}
\left[\int^{M^\star_0}_0 \phi^\star_0 \left(\frac{M^\star}{M^\star_0}\right)^\beta \left(\frac{M}{M^\star}\right)^\alpha e^{-(M/M^\star)} d(\log M^\star)\right] dM &=& \left[-\int^{M/M^\star_0}_\infty \phi^\star_0 \left(\frac{M^\star}{M^\star_0}\right)^\beta \left(\frac{M}{M}\right)^\beta \left(\frac{M}{M^\star}\right)^\alpha e^{-(M/M^\star)} t^{-1} dt\right] d\left(\frac{M}{M^\star_0}\right) \ \ \ \ \ \\
&=& \left[\int^\infty_{M/M^\star_0} \phi^\star_0 \left(\frac{M}{M^\star_0}\right)^\beta t^{-\beta} t^\alpha e^{-t} t^{-1} dt\right] d\left(\frac{M}{M^\star_0}\right)\\
&=& \left[\phi^\star_0 \left(\frac{M}{M^\star_0}\right)^\beta \int^\infty_{M/M^\star_0} t^{\alpha-\beta-1} e^{-t} dt\right] d\left(\frac{M}{M^\star_0}\right)
\end{eqnarray}
Which can be evaluated using the upper incomplete gamma function to give:
\begin{equation}\label{eqn6}
\left[\int\phi(M,M^\star) d(\log M^\star)\right]dM = \phi^\star_0 \left(\frac{M}{M^\star_0}\right)^\beta \Gamma_{M/M^\star_0} (\alpha - \beta) d\left(\frac{M}{M^\star_0}\right)
\end{equation}

\begin{figure}
\epsscale{0.65}
\plotone{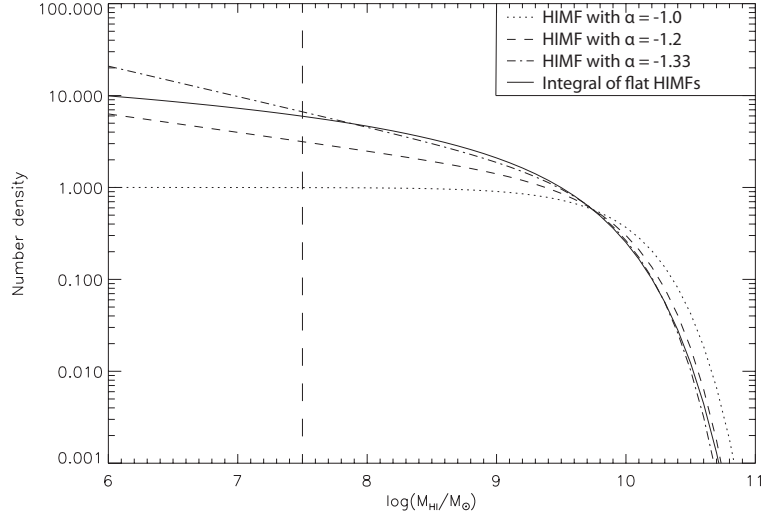}
\caption{Integral of the flat HIMFs (solid line) compared to HIMFs with $\alpha = -1.0$ (dotted), $-1.2$ (dashed) and $-1.33$ (dot-dashed), the last being the same as the ALFALFA field HIMF (\citealt{martin2010}). It can be seen that there is a relatively close match between the integral of the flat HIMFs and the ALFALFA field HIMF for $M > 10^{7.5} M_\odot$ (the limit of the catalogs used in \citealt{jones2016}, indicated here by the long-dashed vertical line), although it diverges below this.}
\label{fig1}
\end{figure}

Equation \ref{eqn6} was evaluated numerically using WolframAlpha\footnote{http://wolframalpha.com} and compared to various values for a single Schechter function, as shown in Figure \ref{fig1}. It can be seen from this that the integral of the flat HIMFs matches the ALFALFA 40\% catalog field HIMF \citep{martin2010}, which is somewhat steeper than the HIMFs found by \citet{jones2016} from the 70\% catalog, relatively well over the range of values (around $10^{7.5}-10^{10.5} M_\odot$) covered by \citeauthor{jones2016}. That there are differences between the Schechter function and my integrated function implies that it would be possible to distinguish between them with data that was either deep enough or had sufficiently small errors, or to determine (which is most likely in the opinion of this author) that both are approximations to a more complex form.

\section{Comparison with data and discussion}

\begin{figure}
\plottwo{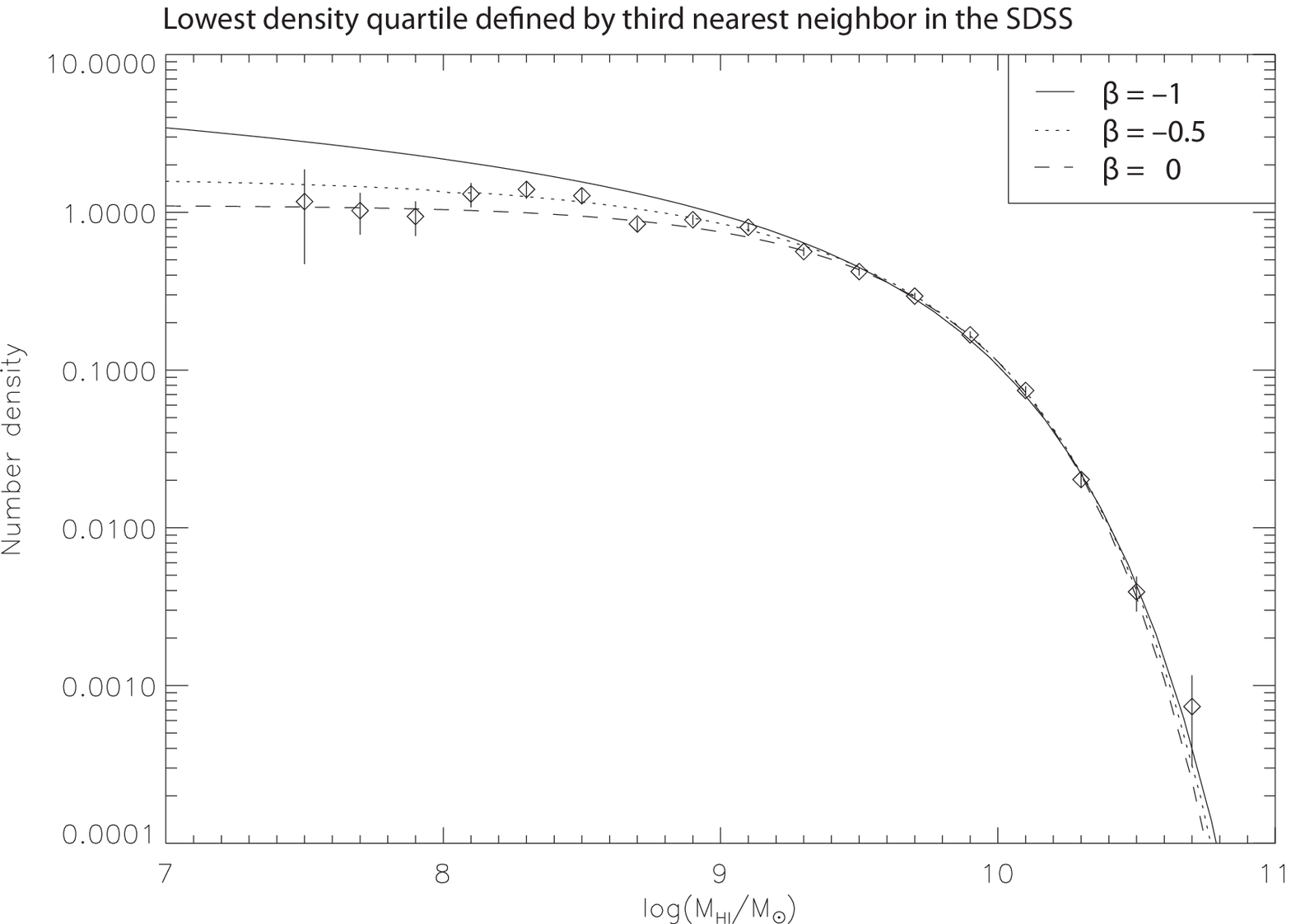}{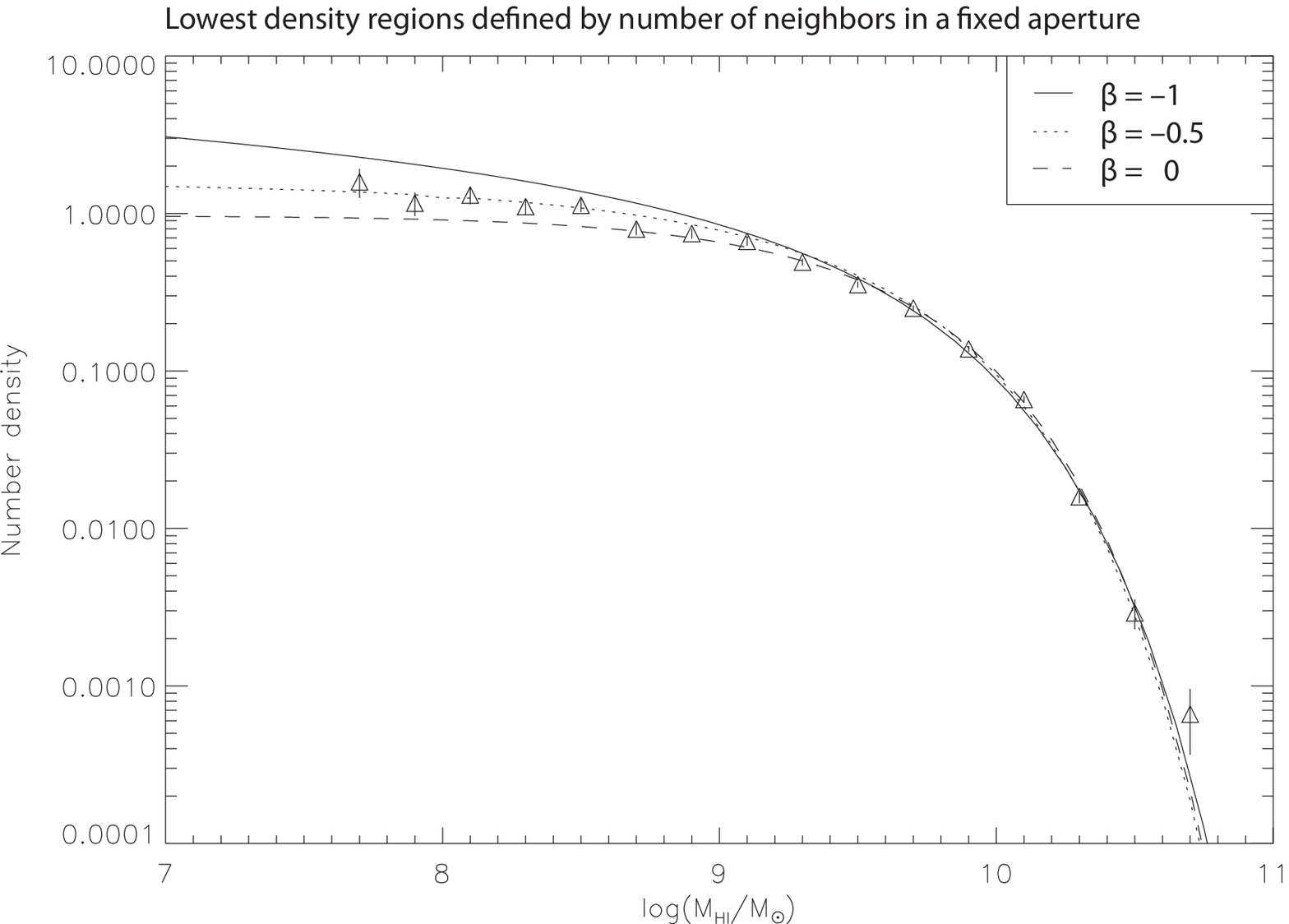}
\plottwo{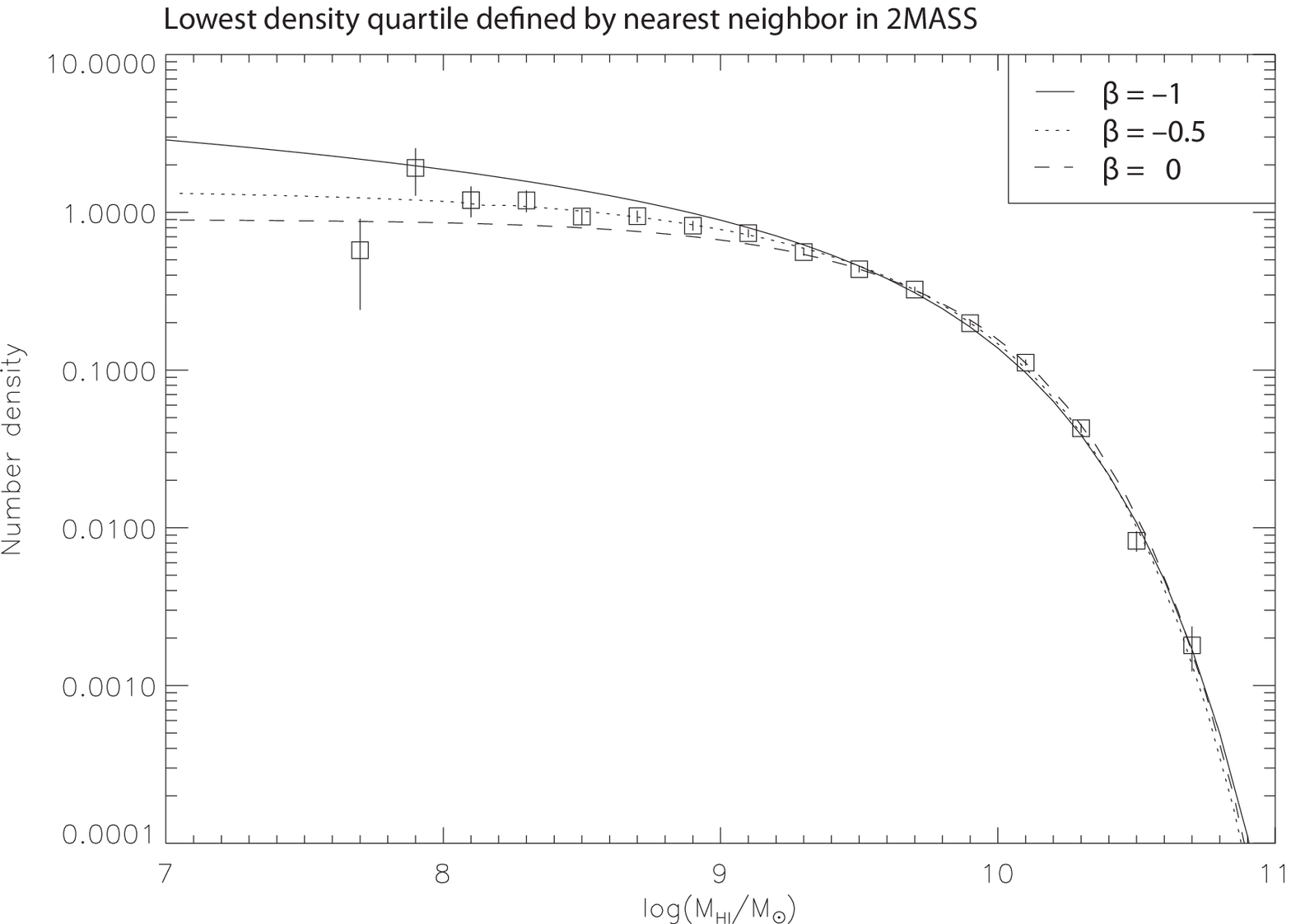}{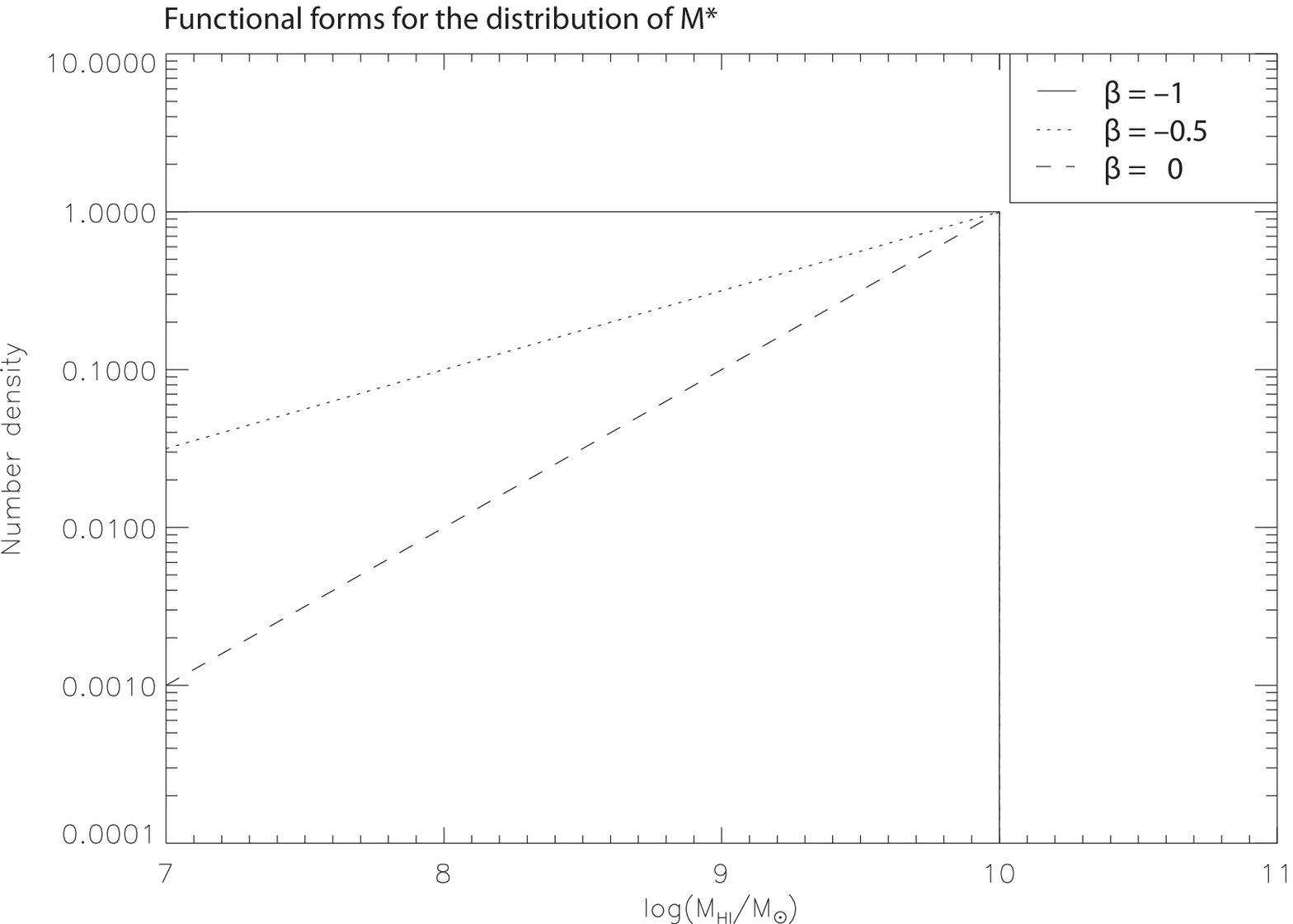}
\caption{Comparison with data for $\beta = -1$ (solid line), $-0.5$ (dotted line) and $0$ (dashed line). Top left: comparison with the HIMF in the lowest density quartile defined by the third nearest neighbor in the SDSS (\citeauthor{jones2016}  Fig. 9); top right: lowest density regions defined by the number of neighbors in a fixed aperture (\citeauthor{jones2016}  Fig. 10); bottom left: the lowest density quartile defined by the nearest neighbor in 2MASS (\citeauthor{jones2016} Fig. 11)' bottom right: functional forms used for the distribution of $M^\star$. Vertical normalization is arbitrary in all plots.}
\label{fig2}
\end{figure}

The model was tested by comparing it with data from the lowest density bins of the various methods of calculation used in \citet{jones2016}: third nearest neighbor in the SDSS, number of neighbors in a fixed aperture, and nearest neighbor in 2MASS. As the model is inspired by the distribution around isolated galaxies, this is the most appropriate density with which to make the comparison, but the shape of the HIMF is similar in all of \citeauthor{jones2016}'s density bins. The integrated HIMF was calculated for $\beta = -1, -0.5$ and $0$ to investigate the effect of changing the shape of the distribution of $M^\star$. Results are plotted in Figure \ref{fig2}, which also shows the function forms of the distribution of $M^\star$ used.

Obviously this is a very simple model, with the distribution of $M^\star$ being at best a first-order approximation to reality. As $\beta = -1$ is a fairly good match (Figure \ref{fig1}) for the 40\% ALFALFA HIMF of Martin et al. (2010), which has a steeper faint-end slope than the \citet{jones2016} HIMFs (based on the ALFALFA 70\% catalog), it is unsurprising that this is slightly higher than the data in Figure \ref{fig2}. However, there is a surprisingly good match between the model and the data for $\beta = -0.5$. $\beta = 0$ falls mostly beneath the data at the faint end (or is too high at the knee, depending on normalization).

The good match for $\beta = -0.5$ demonstrates that the flat local HIMFs, as seen around isolated galaxies and in groups, can sum to give a non-flat global HIMF, as observed by \citet{jones2016}, thus the results of \citeauthor{jones2016} are not necessarily in tension with previous results targeting specific structures. This implies that caution should be used in interpreting the results of both global and local HIMFs as they are probing different aspects of the galaxy population: it is not possible to generalize from local HIMFs to the global population, nor to use the global HIMFs to examine effects on the local scale.

More generally, while this paper assumes a fixed faint-end slope for the purpose of making the analysis tractable, it is possible to speculate that in the real world both the faint-end slope and the position of the knee in the local HIMF could vary with environment even while the global HIMF maintains a constant or near-constant faint-end slope and sees only small changes in the position of the knee. It is trivial, for example, to see that if $\alpha = -1.2$ and $\beta = 0$, the results would simply be a Schechter function with a slope of -1.2. This implies that intermediate values of $\alpha$ and $\beta$ between this and the $\alpha = -1$, $\beta = -0.5$ considered earlier would also give similarly-shaped functions, making it quite possible for the combination of $\alpha$ and $\beta$ to vary with environment while maintaining a global HIMF that does not change significantly with density.

\section{Conclusion}
It has been demonstrated that it is possible, with a simple model, to reconcile the apparently contradictory findings of a relatively steep global faint-end slope to the HI mass function that is constant across different galactic densities, as seen by \citet{jones2016} and observations of a flat local slope to the HI mass function in galaxy groups and around isolated galaxies. This points to the possibility of a `conspiracy' causing the differing local HIMFs seen at different galactic densities to always sum to form similar global HIMFs.

\acknowledgements
The Arecibo Observatory is operated by SRI International under a cooperative agreement with the National Science Foundation (AST-1100968), and in alliance with Ana G. Méndez-Universidad Metropolitana, and the Universities Space Research Association. RFM wishes to thank Andrew Seymour, Rhys Taylor, Olivia Keenan and members of the ALFALFA team for useful discussions, and to thank Mike Jones for making his data available. I also 
thank the anonymous referee whose comments helped greatly in clarifying and improving this paper.

\end{document}